\documentclass[10pt,twocolumn,eqsecnum,floats,showpacs,nofootinbib,amsmath,amssymb,aps,prd]{revtex4-1}

\usepackage{amsmath,amssymb}
\usepackage{hyperref}

\begin{document}

\title{Uniqueness of the Fock quantization of scalar fields under mode preserving canonical transformations varying in time}

\author{Jer\'onimo Cortez}
\email{jacq@ciencias.unam.mx}
\affiliation{Departamento de F\'\i sica, Facultad de Ciencias, Universidad Nacional Aut\'onoma de
M\'exico, M\'exico D.F. 04510, Mexico.}

\author{Luc\'{\i}a Fonseca}
\email{lfonseca@ucm.es}
\affiliation{Facultad de Ciencias F\'{\i}sicas, Universidad Complutense de Madrid, Ciudad Universitaria, 28040 Madrid, Spain}

\author{Daniel \surname{Mart\'{\i}n-de~Blas}}
\email{daniel.martin@iem.cfmac.csic.es}

\author{Guillermo A. Mena~Marug\'an}
\email{mena@iem.cfmac.csic.es}
\affiliation{Instituto de Estructura de la Materia, IEM-CSIC, Serrano 121, 28006 Madrid, Spain}

\begin{abstract}

We study the Fock quantization of scalar fields of Klein-Gordon type in nonstationary scenarios propagating in spacetimes with compact spatial sections, allowing for different field descriptions that are related by means of certain nonlocal linear canonical transformations that depend on time. More specifically, we consider transformations that do not mix eigenmodes of the Laplace-Beltrami operator, which are supposed to be dynamically decoupled. In addition, we assume that the canonical transformations admit an asymptotic expansion for large eigenvalues (in norm) of the Laplace-Beltrami operator in the form of a series of half integer powers. Canonical transformations of this kind are found in the study of scalar perturbations in inflationary cosmologies, relating for instance the physical degrees of freedom of these perturbations after gauge fixing with gauge invariant canonical pairs of Bardeen quantities. We characterize all possible transformations of this type and show that, independently of the 
initial field description, the combined criterion of requiring (i) invariance of the vacuum under the spatial symmetries and (ii) a unitary implementation of the dynamics, leads to a unique equivalence class of Fock quantizations, all of them related by unitary transformations. This conclusion provides even further robustness to the validity of the proposed criterion, completing the results that have already appeared in the literature about the uniqueness of the Fock quantization under changes of field description when one permits exclusively local time dependent canonical transformations that scale the field configuration.
\end{abstract}


\maketitle

\section{Introduction}
\label{s1}

It is well known that the construction of a quantum theory to describe a system which is totally characterized classically is a process plagued with ambiguities, with choices at various steps of the quantization procedure that lead to different physical predictions. Even if the classical system is identified with an algebra of functions on phase space, constructed out of the set of elementary variables that one had selected, distinct nonequivalent quantum representations are generally possible. For simple systems, a unique representation can be picked out by imposing certain requirements, usually related to symmetries of the system or to a good physical (or mathematical) behavior of the representation. For instance, in Quantum Mechanics, when only a finite number of degrees of freedoms are present, one can determine a unique representation of the Weyl algebra (formed with the exponentials of the position and momentum variables, multiplied by imaginary numbers) by demanding irreducibility, unitarity, and 
strong continuity of the representation \cite{simon}. For fieldlike systems, the existence of an infinite number of degrees of freedom makes the situation much more intricate. This happens even for fields with linear evolution equations for which, after writing the canonical commutation relations (CCR's) in a way similar to that of the Weyl algebra, one can adopt a representation of the Fock type \cite{wald}, where one has a notion of vacuum and particle at hand (not necessarily well founded from a physical viewpoint, but rather auxiliary in many situations). Generically, there exist infinitely many Fock representations of the CCR's that are not equivalent under unitary transformations \cite{wald}. In these circumstances, one must call for additional criteria to select a unique representation \cite{criteria}, up to unitary transformations which do not affect the physical content. When the field propagates in a highly symmetric spacetime, typically, the criterion consists of demanding that all the 
quantization structures incorporate this symmetry, like e.g. with Poincar\'e symmetry in Minkowski spacetime \cite{wald}. But for generic curved spacetimes, no such criterion exists.

The loss of predictability owing to the persistence of ambiguities is particularly relevant in the consideration of cosmological settings, since one can only make observations in the universe in which one lives, instead of performing a(n ideally infinitely) large number of measurements to discern among potential candidates for a quantum theory. Moreover, this problem is exacerbated in cosmology by the obstructions to find windows for the detection of quantum effects. For this kind of cosmological situations, where fields propagate in nonstationary spacetimes, a criterion has been proposed recently to pick out a preferred unique class of Fock quantizations for Klein-Gordon (KG) fields, assuming compact spatial topology.\footnote{The spacetime where the field propagates may correspond to a physical background \cite{wald,b-d}, an effective background which incorporates some types of quantum corrections \cite{LQCap,FMOV,FMO,AKL}, or just an auxiliary background that facilitates the field description \cite{
reGowdy,feasab,unit-gt3}.} This criterion consists in demanding that the vacuum state be invariant under the spatial symmetries of the system, complemented with the additional requirement that the evolution admit a unitary implementation. The criterion was introduced initially in the quantization of Gowdy models \cite{feasab,unit-gt3,unit-gt4,unique-gowdy-1,CMV5,cmsv2}, and then extended to KG fields with time dependent mass defined on the circle \cite{cmsv}, on the three-sphere \cite{CMV8,CMOV-S3S,CMOV}, and finally on spatial manifolds of arbitrary compact topology in three or less dimensions \cite{cmov-gen,Crit}. This includes the physically relevant case of compact sections with three-torus topology \cite{CGMMV}, with applications to realistic models in cosmology, in accordance with current observations of the universe, which favor spatial flatness \cite{wmap}.

In fact, the ambiguities that the proposed criterion has been shown to resolve in the Fock quantization are so far of two kinds \cite{cmov-gen,Crit}. On the one hand, once a canonical pair is chosen for the scalar field, one finds the well known ambiguity that we have commented about the choice of a Fock representation of the CCR's. In more detail, let us explain that the relevant information for the selection of a Fock representation is captured in the choice of a complex structure (CS). A CS is a real linear map $J$ on phase space whose square is minus the identity and which leaves invariant the symplectic structure, i.e., it is a symplectomorphism. Recall that the symplectic structure $\Omega(\cdot,\cdot)$ is a bilinear map on phase space \cite{wald} that determines the CCR's. \footnote{Loosely speaking, the symplectic structure encodes the information about the canonical Poisson brackets.} For the construction of a Fock representation, one also requires that the CS be compatible with the symplectic 
structure, in the sense that the combined map $\Omega(J\cdot,\cdot)$ be positive definite. This condition allows one to define an inner product on phase space, given by  $[\Omega(J\cdot,\cdot)-i\Omega(\cdot,\cdot)]/2$. Starting then from the positive frequency sector on this space, obtained with the projector $(1-i J)/2$, and completing it with the introduced inner product, one arrives at the one-particle Hilbert space of the theory. The Fock space is the direct sum of the symmetric tensor products of this one-particle Hilbert space. The proposed criterion removes the ambiguity inherent to the choice of representation, selecting the class of unitary equivalence which includes the CS $J_0$ that would be naturally associated with the case of a massless KG field.

On the other hand, in nonstationary scenarios, it is natural to consider different field descriptions that are related by a scaling of the field by a time dependent function. This scaling can be seen as part of a linear canonical transformation, in which the field momentum gets the inverse scaling and, in addition, a contribution of the field configuration can be added to the momentum, with all the coefficients of the linear transformation  allowed to vary in time. This type of canonical transformation is local, and changes the field dynamics, given its time dependence. These scalings are often found in cosmology (see e.g. Ref. \cite{mukhanov} for the typical scaling of cosmological perturbations), and absorb part of the time variation of the field, assigning it to the spacetime background. The subsequent modification of the field dynamics explains why the criterion of unitary implementation of the evolution has different consequences in the distinct field descriptions obtained in this manner. The result is 
that, together with the invariance under the spatial symmetries, the requirement of unitary evolution picks out a unique canonical pair for the system among all those related by this family of local, time dependent linear canonical transformations \cite{Crit}.

For the discussion of these two types of ambiguities and the application of the uniqueness results when the proposed criterion is imposed, the starting point is that the field system admits, with an appropriate choice of time and after a suitable scaling, a formulation in terms of a KG field $\varphi$ with a time dependent mass $s(t)$ that propagates in a(n auxiliary) ultrastatic spacetime.\footnote{Some mild conditions are imposed on the mass function $s(t)$ to reach the results of Refs. \cite{cmov-gen,Crit}. It suffices that this function is twice differentiable in the considered time domain, with a second derivative that is integrable in every compact time subinterval.} Namely, the field equations can be written in the form
\begin{equation}
\label{infeq}
\ddot\varphi-\Delta\varphi+s(t)\varphi=0,
\end{equation}
together with a Hamiltonian equation for the momentum $p_{\varphi}$ given by $p_{\varphi}=\sqrt{h} \dot{\varphi}$. Here, the dot stands for the time derivative and $\Delta$ denotes the standard Laplace-Beltrami (LB) operator associated with the spatial metric $h_{ij}$ of the background. The determinant of this spatial metric is called $h$. The corresponding spatial sections are assumed to have compact topology, and dimension equal to $d$. When $d\le3$, as we have mentioned, the criterion of a unitary implementation of the evolution and of invariance under the spatial symmetries selects the class of unitary equivalence that includes $J_0$ as the only valid CS's \cite{cmov-gen}, and eliminates the ambiguity concerning the choice of field description as far as time dependent scalings of the field configuration and linear redefinitions of the field momentum are concerned \cite{Crit}.

In the context of cosmological perturbations in an inflationary scenario, it has also been proven that the above criterion guarantees the uniqueness of the Fock quantization even if the field equations are modified with certain kinds of subdominant terms. More explicitly, expanding the field configuration and momentum in eigenmodes of the LB operator, the KG field equations with time varying mass may present corrections that depend on the considered mode, but that vanish sufficiently fast in the ultraviolet limit, i.e., for large eigenvalues (in norm) of the LB operator. These corrections include mode dependent subdominant contributions, on the one hand to the time dependent mass and, on the other hand, in the form of a damping term \cite{FMOV}. In fact, the damping terms can be absorbed by means of a suitable canonical transformation, which is nonetheless nonlocal \cite{FMOV}. The presence of these corrections does not spoil the ability of the proposed criterion to pick out a unique Fock quantization, 
showing that the results go beyond the systems which admit a formulation precisely as a KG field in an ultrastatic spacetime.

An important limitation of these uniqueness theorems is that they only cover changes of field descriptions related by local (linear) time dependent canonical transformations. In practical situations, at least in cosmological settings, one decompose the field in eigenmodes of the LB operator, taking full advantage of the fact that these modes decouple dynamically, and then introduces a CS by defining annihilation and creationlike variables for each mode. This definition is often time dependent because, in nonstationary scenarios, the frequencies of the modes change. In order to relate the choice of annihilation and creationlike variables with that determined by the CS $J_0$ in the field description in which \eqref{infeq} applies, one needs to introduce a linear canonical transformation that, in general, depends on the considered mode. Besides, as we have said, the transformation is typically time dependent. What is no more true is that the transformation is local, and in particular that it amounts to a global 
scaling of the field when configurations are considered. In order to analyze the extent to which one can guarantee the uniqueness of the Fock quantum theory, and discuss the validity of the criterion that we are putting forward to deal as well with this generalized framework and remove the subsequent ambiguity, we will study in this work the effect in the quantization of canonical transformations of the mentioned type. In doing so, we will allow for the most general linear canonical transformation of that kind, without restricting the changes of field configuration to be contact transformations that depend on time (and on the considered mode) but, on the contrary, permitting also the inclusion of contributions of the momentum modes. Let us emphasize that, since these time varying linear canonical transformations are mode dependent, they are generally nonlocal: inverse powers of the LB eigenvalue are in fact obtained via the inversion of the LB operator, which is a nonlocal operation.

Moreover, in principle it is not clear if, starting with a field description in which Eq. \eqref{infeq} is satisfied, one can reach another description with a similar field equation (maybe up to subdominant terms in the ultraviolet regime for the mass and the damping) by means of a mode dependent canonical transformation of this sort. If this is possible, one would find a potential tension in the real implications of the results about the uniqueness of the choice of Fock quantization, namely, those results would provide a privileged Fock quantization for each of the descriptions with field equations of the form \eqref{infeq}, but their relation being provided by nonlocal transformations, we could not assure that such distinct quantizations are all unitarily equivalent. If they were not, a new ambiguity would arise in the selection of a Fock formulation.

Actually, a situation of this kind is found in the study of scalar perturbations around Friedmann-Robertson-Walker (FRW) spacetimes. These perturbations provide seeds for the large scale structure and explain the origin of the cosmic microwave background \cite{lif,mukhanov,mukhanov1}. In order to circumvent the problems posed by gauge transformations, an approach which is sometimes followed is to adopt a gauge, e.g. the longitudinal one, and express the physical degrees of freedom in terms of quantities defined in the system with that gauge fixation. Knowing the effect of the gauge transformations, these quantities can be reexpressed in any other gauge. Another approach is to directly eliminate any dependence on the choice of gauge by considering gauge invariant quantities, such as the Bardeen potentials \cite{bardeen}. In Ref. \cite{FMOV}, it was proven that a canonical pair for the description of the scalar cosmological perturbations was the one formed by the energy density and the matter velocity 
perturbations, which are Bardeen potentials \cite{bardeen}. In that same work, it was shown that the change from the canonical pair that describes the perturbations in the longitudinal gauge to this pair of gauge invariants is given in fact by a time dependent linear canonical transformation which is mode dependent \cite{FMOV}, and hence of the form that we are going to analyze in the present article. For the scalar perturbations, and up to subdominant terms in the ultraviolet, the field equations turned out to be of the KG type under discussion, both for the canonical pair chosen in the longitudinal gauge and for the pair of Bardeen potentials. In that particular case, it was proven also in Ref. \cite{FMOV} that the two privileged Fock quantizations chosen by the criterion of spatial symmetry invariance and unitary evolution when studying, respectively, the gauge fixed system and the gauge invariants, happened to be unitarily equivalent, preserving in this way the uniqueness of the quantum theory, up to 
unitary transformations. Nonetheless, it was far from clear whether this was a result specific of the considered change or not, and why a mode dependent canonical transformation compatible with the mentioned criterion, like the one encountered in the case of scalar perturbations, could exist. In the light of the present work, these conclusions for the model of scalar perturbations in cosmology are just a particular application of more general results. The example provided by this model is explained in detail in the Appendix.

With these motivations, our work has a twofold aim. On the one hand, we want to characterize the most general mode and time dependent linear canonical transformation that relates different field descriptions of the system where a unitary implementation of the evolution is possible, while respecting the invariance under the spatial symmetries. Note that it is for each of these descriptions that the criterion that we put forward selects, respectively, a privileged Fock quantization. On the other hand, we will study then whether these privileged quantizations are equivalent or not, by elucidating whether the canonical transformation in question admits or not a unitary implementation. Our analysis will be restricted to canonical transformations which preserve the mode decomposition of the field in eigenfunctions of the LB operator, i.e., to transformations which respect the dynamical decoupling of these modes and do not mix them. In addition, we will assume that the relations that provide the canonical 
transformation admit an asymptotic expansion in the ultraviolet sector in the form of a Laurent series of the square root of (minus the) eigenvalue of the LB operator.

At this point of the discussion, it is worth recalling that a linear canonical transformation $T$ admits a unitary implementation in the representation determined by a CS $J$ if and only if the antilinear part of $T$, given by $(T+JTJ)/2$, is a Hilbert-Schmidt operator \cite{shale}, i.e., the product of this antilinear part by its adjoint has a finite trace. This condition is equivalent to the square summability of the beta Bogoliubov coefficients.  Recall also that these Bogoliubov coefficients relate the annihilation and creationlike variables of the Fock representation with their images under the considered transformation, and that they are usually called alpha and beta coefficients, depending on whether they correspond to the linear or the antilinear part of the transformation, respectively. Finally, let us remark that, from a physical viewpoint, the square summability of the beta coefficients is simply the condition that the particle production under the analyzed transformation be finite, namely, that 
the transformed vacuum have a finite number of particles, if one employs the particle concept associated with the original vacuum.

The rest of the paper is organized as follows. In Sec. \ref{s2} we will describe the system under study. The kind of nonlocal, time dependent linear canonical transformations that we want to analyze will be introduced in Sec. \ref{s2b}, investigating whether they can relate descriptions with field equations of the KG type \eqref{infeq}, up to certain subdominant terms in the ultraviolet regime. In Sec. \ref{s3} we will completely characterize the most general canonical transformation with the desired properties, and discuss its unitary implementation in the original Fock quantum theory. In addition, we will obtain the form of the subfamily of canonical transformations that lead to KG field equations without mode dependent corrections to the mass term. The conclusions will be presented in Sec. \ref{s5}. Finally, an Appendix is included.

\section{The Klein-Gordon field}
\label{s2}

Our starting point is a real KG field $\varphi$ with a time dependent mass, propagating in an ultrastatic spacetime which is globally hyperbolic, of the form $\mathbb{I}\times\Sigma$, where $\mathbb{I}$ is a connected and not necessarily unbounded time interval (if $\mathbb{I}$ is the union of several connected components, one can restrict the discussion just to one of those components) and $\Sigma$ is a spatial manifold of compact topology. The spatial sections, isomorphic to $\Sigma$, are equipped with the metric $h_{ij}$. The field satisfies the dynamical equation \eqref{infeq}. The dynamical evolution is completed with the Hamiltonian equation that identifies the field momentum $p_{\varphi}$ with the time derivative of the field, densitized by a factor of $\sqrt{h}$. Notice that the system is not stationary, owing to the time dependence of the mass, which can also be interpreted as a quadratic potential term that varies in time. This kind of KG equation can be obtained, e.g., from those of scalar fields 
in nonstationary spacetimes by means of a scaling of the field \cite{CMOV-S3S}, as it is the case of test KG fields of constant mass in FRW spacetimes.

The phase space can be obtained from the Cauchy data at an arbitrary time $t_{0}\in\mathbb{I}$: $(\varphi,p_{\varphi})=(\varphi,\sqrt{h}\dot{\varphi})_{|_{t_{o}}}$. The symplectic structure is that corresponding to the standard Poisson brackets $\{\varphi(x),p_{\varphi}(y)\}=\delta(x-y)$, where $\delta(x)$ is the Dirac delta on $\Sigma$.

The criterion of invariance under the spatial symmetries of the field equations \footnote{One can take as the group $G$ of such symmetries the maximal subgroup of the unitary group of transformations which commute with the LB operator. One can as well identify $G$ with a subgroup of the former maximal subgroup, provided that all its corresponding irreducible representations differ \cite{Crit}.} and of a unitary dynamics selects a unique class of unitary equivalence of CS's for the Fock representation of the KG field \cite{cmov-gen}. This class of CS's is the one which includes $J_0$, namely, the CS that one would naturally associate to the case of a massless field:
\begin{equation}
\label{fmsfcs}
J_{0}\left(\begin{array}{c}
\varphi\\
p_{\varphi}
\end{array}\right)=\left(\begin{array}{cc}
0 & -(-h\Delta)^{-\frac{1}{2}}\\
(-h\Delta)^{\frac{1}{2}} & 0
\end{array}\right)\left(\begin{array}{c}
\varphi\\
p_{\varphi}
\end{array}\right).
\end{equation}
Since $J_{0}$ is constructed out of the LB operator, it is invariant under the spatial symmetries and leads to a Fock representation with the same property.

For convenience, we decompose the canonical pair of field configuration and momentum in a mode expansion using a basis of eigenfunctions of the LB operator (in the Hilbert space of square integrable functions on $\Sigma$ with the volume element determined by the spatial metric). We will call $q_{nl}$ the coefficients of the real modes in the expansion of the field configuration $\varphi$, and $p_{nl}$ the corresponding coefficients for the field momentum. The subindex $n$ is a positive integer that designates the eigenspaces of the LB operator with different eigenvalue, $-\omega_n^2$, according to the ordering in the increasing sequence of positive numbers $\{\omega_n\}$, so that $\omega_n> \omega_{n'}$ if and only if $n> n'$. Note that the compactness of the topology guarantees that the sequence is discrete, and also that the sequence tends always to infinity when $n\rightarrow \infty$. On the other hand, we suppose that the zero modes of our field have already been removed, and quantized by suitable 
methods in Quantum Mechanics, so that we can ignore them in the following. Let us emphasize that the removal of a finite number of degrees of freedom does not modify the field behavior of the system, nor the issues related to the unitary implementation of the dynamics owing to the existence of an infinite number of modes in the ultraviolet sector (large $\omega_n$). The degeneration $\mathfrak{g}_n$ of each of the eigenspaces of the LB operator is taken into account by the subindex $l$. For a fixed value of $n$, the index $l$ can only take a finite number of positive integer values, $l=1,\ldots,\mathfrak{g}_{n}$. Finally, the Hamiltonian equation for the time derivative of the field configuration translates, after the mode expansion, into the equation $p_{nl}=\dot{q}_{nl}$.

Substituting the mode expansion of the field in Eq. \eqref{infeq}, one gets that the real modes $q_{nl}$ satisfy the equations of motion
\begin{equation}
 \label{inrmeqm}
 \ddot{q}_{nl}+[\omega^{2}_{n}+s(t)]q_{nl}=0.
\end{equation}
Clearly, the different modes decouple in the dynamics. It is also obvious that this equation of motion is independent of the label $l$ and therefore all the modes in the same eigenspace of the LB operator have the same dynamics. For the sake of simplicity in the notation then, from now on, we will obviate the degeneration label $l$ in our equations, unless it is necessary to avoid confusions.

\section{Time and mode dependent canonical transformations}
\label{s2b}

As we have already mentioned in Sec. \ref{s1}, mode dependent linear canonical transformation that vary in time appear naturally in Quantum Field Theory in nonstationary spacetimes, and in particular in its application to cosmological perturbations. For instance, in the Appendix we discuss in detail the transformation of this kind that relates the canonical pair of degrees of freedom that describe scalar perturbations around a closed FRW universe in longitudinal gauge and the canonical pair of Bardeen potentials which provide a gauge invariant formulation \cite{bardeen,FMOV}.

We are interested only in nonlocal canonical transformations that lead from a field description with equations of motion of the form  \eqref{inrmeqm} to another field description whose modes satisfy a similar equation, but now with a possibly different time dependent mass, which may include contributions that depend on the considered mode but become subdominant in the ultraviolet regime, i.e. when $\omega_{n}\rightarrow \infty$. Specifically, we assume that the new mass is given by
\begin{equation}
\label{nmt}
M_{n}(t)=\tilde{s}(t)+\mathcal{O}\left(\omega^{-1}_{n}\right),
\end{equation}
where $\tilde{s}(t)$ is a mode independent time function which provides the ultraviolet limit. The criterion of spatial symmetry invariance and unitary evolution can be applied to select a unique class of Fock representations for fields that satisfy a KG equation with a mass term of the above form \cite{FMOV}.

Let us now discuss what kind of nonlocal and time dependent linear canonical transformation allows one to pass from the original canonical pairs of modes $(q_{n},p_{n})$, subject to the equations of motion \eqref{inrmeqm} and $p_{n}=\dot{q}_{n}$, to a new set of canonical modes $(Q_{n},P_{n})$, satisfying the same equations of motion, but now with a mass term supplied by $M_{n}(t)$ instead of $s(t)$. As we have mentioned, we will only consider canonical transformations that do not mix modes (respecting their dynamical decoupling). Therefore, the most general transformation that we are going to study is
\begin{eqnarray}
\label{cmdt}
Q_{n}&=&f_n(t)p_{n}+g_{n}(t)q_{n}, \\
P_{n}&=&\frac{1}{g_{n}(t)}\left[1-f_{n}(t)B_{n}(t)\right]p_{n}-B_{n}(t)q_{n},
\label{cmdt2}
\end{eqnarray}
where $f_{n}(t)$, $g_{n}(t)$, and $B_{n}(t)$ are mode and time dependent real functions. Nonetheless, $B_n(t)$ is completely determined by $f_{n}(t)$ and $g_{n}(t)$, via the imposition of the Hamiltonian equations $P_{n}=\dot{Q}_{n}$. Thus, the canonical transformations are characterized just by the functions $f_{n}(t)$ and $g_{n}(t)$. A simple calculation, employing the dynamical equations of the original canonical pair and the transformation maps \eqref{cmdt} and \eqref{cmdt2}, shows that the Hamiltonian equation for $\dot{Q}_{n}$ implies that
\begin{eqnarray}
\nonumber
B_{n}(t)&=&[\omega^{2}_{n}+s(t)]f_{n}(t)-\dot{g}_{n}(t)\\
&=&\frac{1-\dot{f}_{n}(t)g_{n}(t)-g_{n}^{2}(t)}{f_{n}(t)}.
\label{c1}
\end{eqnarray}
This expression not only determines $B_{n}(t)$ in terms of $f_{n}(t)$ and $g_{n}(t)$, but also imposes a condition on these two functions, given by the equality between the right terms in both lines. An additional condition comes from the generalized KG equation for $Q_{n}$. It is not difficult to see (using again the form of the canonical transformation and the original field equations) that the only nontrivial requirement derived from this condition refers to the modification of the mass, which is given by
\begin{eqnarray}
\nonumber
M_{n}(t)-s(t)&=&\frac{f_{n}(t)\dot{s}(t)+2\dot{f}_{n}(t)[\omega^{2}_{n}+s(t)]-\ddot{g}_{n}(t)}{g_{n}(t)}\\
&=& -\frac{\ddot{f}_{n}(t)+2\dot{g}_{n}(t)}{f_{n}(t)}.
\label{c2}
\end{eqnarray}
In addition to the relation between the new and the original mass functions, $M_{n}(t)$ and $s(t)$, again, the equality between the right terms of both lines implies another condition on the functions $f_{n}(t)$ and $g_{n}(t)$. In fact, one can easily check that this condition is not functionally independent of the one derived from Eq. \eqref{c1}: the former can be obtained from the latter by differentiation. Therefore, in the following we will only consider the condition arising from formula \eqref{c1}. We can rewrite it in the form
\begin{eqnarray}
\nonumber
& \dot{f}_{n}(t)g_{n}(t)- \dot{g}_{n}(t)f_{n}(t) \quad \quad \quad   \\ & \ + g_{n}^{2}(t)+[\omega^{2}_{n}+s(t)]f_{n}^{2}(t)=1.
\label{stc}
\end{eqnarray}

As our first result, we are going to prove that it is impossible that a canonical transformation of the considered type but limited to be mode independent --and hence local-- can relate two descriptions with dynamical equations of the specific form in which we are interested. Suppose for a moment that $f_{n}(t)=f(t)$ and $g_{n}(t)=g(t)$ are mode independent. It is easy to see then that the only manner in which Eq. \eqref{stc} may be satisfied is that $f(t)=0$, for all $t\in\mathbb{I}$, because otherwise the only mode dependent contribution in that equation, $\omega^{2}_{n}f^{2}(t)$, will not cancel. Now, if $f(t)$ is the zero function, it follows from our equation that $g(t)=\pm1$ at all times, obtaining a trivial canonical transformation that leads to the original field equations and to the same canonical pair (up to a global sign).  Therefore, we are forced to consider functions $f_{n}(t)$ and $g_{n}(t)$ that are mode (and time) dependent. Note that this result is in complete agreement with that about the 
uniqueness of the field description selected by our criterion when only local time dependent canonical transformations are considered \cite{Crit} (reducing just to a scaling, as far as the field configuration is concerned). In Sec. \ref{s3-2}, we will further study restriction \eqref{stc}, as well as the restrictions arising from the behavior \eqref{nmt} that we have assumed for the mode dependence of the new mass term.

\section{Unitary implementation and asymptotic expansion}
\label{s3}

Here, we will study the relationship between the Fock representations determined by the CS $J_0$ in the two field descriptions under discussion: the initial one, with mode variables  $(q_{n},p_{n})$, and the transformed one, with phase space variables $(Q_{n},P_{n})$. We will also elaborate on the condition for the unitary implementation of this canonical transformation, and therefore for the equivalence of the considered Fock quantizations. Assuming the existence of asymptotic expansions --in the form of Laurent series-- in $\omega_n$ when this quantity gets unboundedly large, we will characterize the admissible canonical transformations --i.e., those that satisfy Eq. \eqref{stc} and lead to the desired new mass term behavior-- and show that all of them can be implemented as unitary quantum transformations. We conclude the section by analyzing the case in which the new mass term is mode independent.

\subsection{Beta coefficients of the transformation}
\label{s3-1}

Let us then consider, for the two field descriptions related by the mode dependent canonical transformation, the Fock representation defined by $J_{0}$ \eqref{fmsfcs}, which we know is picked out (or rather its unitary equivalence class) by the criterion of invariance under the spatial symmetries and the unitary implementability of the dynamics. The following are annihilation and creationlike variables associated with the choice of $J_{0}$ in the initial field description (in the sense that the CS has a diagonal action on the basis formed by them)
\begin{equation}
\label{incav}
a_{n}=\frac{1}{\sqrt{2\omega_{n}}}(\omega_{n}q_{n}+ip_{n}), \end{equation}
\begin{equation}
a^{\ast}_{n}=\frac{1}{\sqrt{2\omega_{n}}}(\omega_{n}q_{n}-ip_{n}),
\label{incav2}
\end{equation}
where the symbol $\ast$ denotes complex conjugation. The corresponding variables for the transformed modes are defined in a completely similar way, i.e.,
\begin{equation}
\label{trcav}
b_{n}=\frac{1}{\sqrt{2\omega_{n}}}(\omega_{n}Q_{n}+iP_{n}),
\end{equation}
and $b^{\ast}_{n}$ by the complex conjugate of this equation.

The relation between both sets of annihilation and creationlike variables is given by a Bogoliubov transformation which, recalling that the modes do not mix, has the form
\begin{eqnarray}
b_{n}&=&\alpha_{n}a_{n}+\beta_{n}a^{\ast}_{n}, \\
b^{\ast}_{n}&=&\beta^{\ast}_{n}a_{n}+\alpha^{\ast}_{n}a^{\ast}_{n}.
\end{eqnarray}
Since the transformation is a symplectomorphism (i.e., it preserves the canonical structure), the alpha and beta functions satisfy the relation $|\alpha_{n}|^{2}-|\beta_{n}|^{2}=1$ for all $n\in\mathbb{N}^{+}$. As we have already mentioned, a linear canonical transformation admits a unitary implementation in a given Fock representation if and only if the beta functions of the transformation are square summable. In this case, since we want the canonical transformation to be unitary at all times, the condition that must be satisfied is
\begin{equation}
\label{ssc}
\sum_{n,l}|\beta_{nl}(t)|^{2}=\sum_{n} \mathfrak{g}_{n}|\beta_{n}(t)|^{2} < \infty\quad \forall t\in\mathbb{I},
\end{equation}
where we have taken into account the degeneration $\mathfrak{g}_n$ of the eigenspaces of the LB operator. A direct calculation leads to the expression of the beta coefficients, using in the computations the form of the canonical transformation \eqref{cmdt} and \eqref{cmdt2}, the value of the functions $B_{n}(t)$ [given in the first equality of Eq. \eqref{c1}], the definitions of the two sets of annihilation and creationlike variables, and relation \eqref{stc}. The result is
\begin{equation}
\label{bc}
\beta_{n}(t)=-\frac{\dot{f}_{n}(t)}{2}-\frac{i}{2\omega_{n}}\left[f_{n}(t)s(t)-\dot{g}_{n}(t)\right].
\end{equation}

The asymptotic behavior of the degeneration $\mathfrak{g}_n$ for large $n$ is well known. The number of eigenstates whose eigenvalue is less than a positive number $\bar{\omega}^{2}$ grows in $d$ dimensions at most like $\bar{\omega}^{d}$ \cite{compact}. Therefore if, as in our case, one assumes that the functions $f_n(t)$ and $g_n(t)$ (as well as their derivatives) admit an asymptotic expansion in integer powers of $\omega_n$ when this quantity tends to infinity, so that the same applies to $\beta_n(t)$, the square summability condition for unitary implementation turns out to be satisfied in (all) spatial dimensions $d$ not greater than three if and only if the asymptotic behavior of the beta functions is of the order $\beta_{n} \sim \mathcal{O}\left(\omega_{n}^{-2}\right)$. Considering the real and the imaginary parts of the beta functions independently, we conclude that
\begin{eqnarray}
\label{fguc}
&&\dot{f}_{n}\sim\mathcal{O}(\omega^{-2}_{n}), \\
&&f_{n}(t)s(t)-\dot{g}_{n}(t)\sim\mathcal{O}(\omega^{-1}_{n}).	
\label{fguc2}
\end{eqnarray}
If these two conditions are fulfilled, then the studied Fock quantizations, related by the transformations \eqref{cmdt} and \eqref{cmdt2}, are unitarily equivalent.

\subsection{Unitary implementation of the admissible transformations}
\label{s3-2}

We will now investigate the properties of the admissible canonical transformations, i.e., the mode dependent transformations defined by the sequences of functions $f_{n}(t)$ and $g_{n}(t)$ that satisfy restriction \eqref{stc} (so that the new momentum modes equal the time derivatives of the new configuration modes) and lead to a new mass $M_n(t)$ with the behavior \eqref{nmt}. We will see whether these admissible canonical transformations fulfill the unitarity conditions \eqref{fguc}-\eqref{fguc2}. Throughout our discussion, we will assume that both $f_{n}(t)$ and $g_{n}(t)$ admit a Laurent series expansion in the asymptotic limit $\omega_n\rightarrow\infty$.

Analyzing first restriction \eqref{stc}, it is not difficult to realize that the term $\omega^{2}_{n}f_{n}^{2}(t)$ in that equation can only be compensated in the asymptotic limit of infinitely large $\omega_n$ by other terms if $f_{n}(t)\sim\mathcal{O}(\omega_{n}^{-r})$ with $r\ge1$. To see this, let us divide the equation by $f_{n}^{2}$ and define $G_{n}=g_{n}/f_{n}$, for convenience. The condition can then be rewritten as
\begin{equation}
\label{stc2}
-\dot{G}_{n}+G_{n}^{2}+\omega^{2}_{n}+s(t)=\frac{1}{f^{2}_{n}}.
\end{equation}
Suppose then that we had $r\le0$. The only possibility to compensate the term $G^{2}_{n}+\omega_{n}^{2}$ (notice that $G^{2}_{n}$ is a positive definite function, and $\omega_{n}^{2}$ is always positive as well) would be that $\dot{G}_{n}\sim\mathcal{O}(\omega^{2}_{n})$.\footnote{If $\dot{G}_{n}$ were of asymptotic order $\mathcal{O}(\omega^{N}_{n})$ with $N>2$, then $G^2_n$ would be of order $\mathcal{O}(\omega^{2N}_{n})$ and its contribution could not be balanced. On the other hand, in the case $N<2$, the term $\omega_n^2$ could not be compensated.} But, in that case, $G^{2}_{n}$ would be at least of the order $\mathcal{O}(\omega^{4}_{n})$, and therefore it could not be balanced with any other term in the equation, arriving to a contradiction. Hence, as we anticipated, we necessarily have $f_{n}\sim \mathcal{O}(\omega^{-r}_{n})$ with $r\ge 1$.

One can then straightforwardly see from the dominant term (for large $\omega_n$) in Eq. \eqref{stc2} that $G_{n}\sim f_{n}^{-1}$, and thus $g_{n}\sim \mathcal{O}(1)$. In addition, let us notice that, given the asymptotic behavior allowed for $f_{n}$, condition \eqref{fguc2} for the unitary implementation of the transformation reduces to
\begin{equation}
\dot{g}_{n}\sim\mathcal{O}(\omega_{n}^{-1}).
\label{fguc3}
\end{equation}

Employing our conclusions about the dominant terms of the (Laurent) series in integers powers of $\omega_n$ that provide the asymptotic expansion of the functions $f_n(t)$ and $g_n(t)$, we get that
\begin{eqnarray}
\label{fgf}
f_{n}(t)&=&\frac{c_{r}}{\omega^{r}_{n}}+\sum_{k=r+1}^{\infty}\frac{c_{k}}{\omega_n^{k}}, \\
g_{n}(t)&=&d_{0}+\sum_{k=1}^{\infty}\frac{d_{k}}{\omega^{k}_{n}},
\label{fgf2}
\end{eqnarray}
with $r \geq 1$ and where the $c$'s and $d$'s are time functions (we obviate this time dependence in the notation for simplicity). They are not fully free or independent, for they must satisfy restriction \eqref{stc}. In particular, it is easy to check from that restriction that the functions $d_{k}(t)$ must vanish for $1\le k < r$, because, otherwise, the term $g^{2}_{n}(t)$ would give contributions that could not be compensated in the equation. On the other hand, the necessary and sufficient conditions for a unitary implementation of the transformation (in any possible spatial dimension not greater than three), which are given by Eqs. \eqref{fguc} and \eqref{fguc3}, amount to the requirements $\dot{c}_{1}=\dot{d}_{0}=0$.

We will separate the rest of our analysis in two cases, depending on whether the dominant power $r$ for $f_n(t)$ is equal to the unit or not. We will analyze first the case $r\ge2$. Then, the condition $\dot{c}_{1}=0$ is immediately satisfied, because $c_{1}$ vanishes identically. Besides, the restriction \eqref{stc} existing on the functions of the transformation implies that $d^{2}_{0}=1$, as one can easily see, and consequently one gets that $\dot{d}_{0}$ vanishes indeed. Therefore, when $r\geq 2$, the considered mode dependent canonical transformations can always be implemented as unitary transformations in the original Fock representation determined by $J_0$, and this quantization is unitarily equivalent to the one obtained in the new field description.

Let us consider now the case $r=1$. From the first equality in Eq.  \eqref{c2}, we obtain that the new mass is given by
\begin{eqnarray}
\nonumber
M_{n}(t) & \sim & s(t) - \displaystyle\frac{2\dot{d}_{0}}{c_{1}}\omega_{n} - \displaystyle\frac{\ddot{c}_{1}+2\dot{d}_{1}}{c_{1}}  + 2\displaystyle\frac{c_{2}\dot{d}_{0}}{c^2_{1}}  \\  & &  +\mathcal{O}(\omega^{-1}_{n}).\ \
\end{eqnarray}
Hence, the restriction imposed on $M_{n}(t)$ that it must have a well defined ultraviolet limit [with possible subdominant terms of order $\mathcal{O}(\omega^{-1}_{n})$] requires that $\dot{d_{0}}=0$, and so $d_{0}$ must be a constant. On the other hand, restriction \eqref{stc} on the functions of the transformation imposes at dominant asymptotic order that $d_{0}^{2}+c_{1}^{2}=1$, and thus $c_{1}$ turns out to be a constant function as well, so that in particular we have $\dot{c}_{1}=0$. As a result, we conclude that Eqs. \eqref{fguc} and \eqref{fguc3} are satisfied, and all admissible canonical transformations happen to admit a unitary implementation in the original Fock quantization determined by $J_0$. Therefore, as it occurs as well for the other case $r\geq 2$, the new Fock quantizations are unitarily equivalent to the original one, and our criterion of invariance under the spatial symmetries and of unitary dynamics selects a unique Fock quantum theory for the KG field, up to unitary transformations, 
as we wanted to prove.

\subsection{Characterization of the admissible transformations}
\label{s3-3}

Let us elaborate on the characterization of the admissible canonical transformations; in particular, let us discuss in more detail their asymptotic behavior.

In the case $r=1$, we identify two distinct types of transformations, namely, those with $d_{0} \neq 0$ and those with $d_{0}=0$. A direct inspection shows that when the considered transformations have $d_{0} \neq 0$, the modes of the new field are proportional to the original modes in the ultraviolet regime. Hence, both sets of modes have a similar behavior. On the contrary, for transformations with $d_{0}=0$, the modes of the new field are defined by linear combinations which have a vanishing ultraviolet limit, something which may seem surprising. This is compensated by the behavior of the modes of the new momentum, which blow up when $\omega_n$ goes to infinity. In the Appendix, we will show that, in the context of cosmological perturbations around closed FRW spacetimes, the transformation between the canonical pair for scalar perturbations in the longitudinal gauge and the corresponding pair of Bardeen potentials is precisely a transformation with this behavior \cite{FMOV}. For this kind of 
transformations with vanishing $d_0$, it is easy to check that restriction \eqref{stc} requires that $c_{2}=0$.

In total, when $r=1$, the admissible canonical transformations are determined by (i) functions of the form
\begin{eqnarray}
f_{n}(t)&=&\frac{c_{1}}{\omega_{n}}+\sum_{k=2}^{\infty}\frac{c_{k}(t)}{\omega^{k}_{n}}, \\ g_{n}(t)&=&d_{0}+\sum_{k=1}^{\infty}\frac{d_{k}(t)}{\omega^{k}_{n}},
\end{eqnarray}
with $d_{0}$ and $c_{1}$ being two nonvanishing constants that satisfy $d_{0}^2+c_{1}^{2}=1$, or (ii) by functions of the type
\begin{eqnarray}
\label{pt2}
f_{n}(t)&=&\pm\frac{1}{\omega_{n}}+\sum_{k=3}^{\infty}\frac{c_{k}(t)}{\omega^{k}_{n}}, \\ g_{n}(t)&=&\frac{d_{1}(t)}{\omega_{n}}+\sum_{k=2}^{\infty}\frac{d_{k}(t)}{\omega^{k}_{n}}
\end{eqnarray}
when $d_{0}=0$.

For the sake of clarity, we have made explicit the time dependence of the coefficients in the expansions. In both cases, the asymptotic limit $\omega_n\rightarrow \infty$ of the new mass term $M_n(t)$ is given by $s(t)-2\dot{d}_{1}(t)/c_{1}$, so that, in general, the mass is modified in this limit with respect to its value in the original field description.

Finally, let us return to the case $r\ge2$. In this situation, the admissible canonical transformations are determined by functions
\begin{eqnarray}
f_{n}(t)&=&\frac{c_{r}(t)}{\omega^{r}_{n}}+\sum_{k=r+1}^{\infty}\frac{c_{k}(t)}{\omega^{k}_{n}}, \\
g_{n}(t)&=&\pm 1+\sum_{k=r}^{\infty}\frac{d_{k}(t)}{\omega^{k}_{n}}.
\end{eqnarray}
Therefore, the configuration and momentum modes of the new field description coincide (possibly, up to a sign) with the original ones in the asymptotic limit. On the other hand, the ultraviolet limit of the new mass term $M_n(t)$ is
\begin{equation}
s(t) -\frac{\ddot{c}_{r}(t)+2\dot{d}_{r}(t)}{c_{r}(t)}.
\end{equation}
Again, we see that the functions $M_n(t)$ are not only typically mode dependent, but their asymptotic limit differs from the original mass $s(t)$ in general. In order that the original mass be recovered in the limit of infinitely large $\omega_n$, it is necessary that $d_1$ be a constant if $r=1$, and that $\ddot{c}_{r}=-2\dot{d}_{r}$ in cases with $r\ge2$.

\subsection{Transformations with mode independent mass}
\label{s3-4}

We will now study the subclass of mode and time dependent canonical transformations (of the form considered in Sec. \ref{s3-3}) which lead to a new field description with a mode independent mass, i.e., all the modes of the new field evolve with the same mass term. To emphasize this fact, we will write $M_{n}(t)=M(t)$. Recalling relation \eqref{c2}, we conclude that the functions $g_{n}(t)$ must satisfy the restriction
\begin{equation}
\label{gfs}
g_{n}(t)=-\frac{1}{2}\dot{f}_{n}(t)+\frac{1}{2}\int^t \! S(\bar{t})f_{n}(\bar{t})d\bar{t},
\end{equation}
where $S(t)=s(t)-M(t)$ is the difference between the original and the new mass functions, and a global integration constant has been absorbed, letting unspecified the initial time in the last integration. Thus, the functions $g_{n}(t)$ are completely determined (up to time constants that may depend on the mode) by the functions $f_{n}(t)$ and by the difference between the two considered mass functions. Using the above expression for $g_{n}(t)$, the restriction \eqref{stc} on the functions that determine the canonical transformation becomes
\begin{eqnarray}
\nonumber
&-&\frac{1}{4}\dot{f}_{n}^{2}(t)+\frac{1}{2}\ddot{f}_{n}(t)f_{n}(t)+\frac{1}{4}\left[\int^t \! S(\bar{t})f_{n}(\bar{t})d \bar{t}\right]^{2}    \\  \ & +& [\omega^{2}_{n}+\bar{s}(t)]f_{n}^{2}(t)=1. \ \ \
\label{tdmtfc}
\end{eqnarray}
Here $\bar{s}(t)=[s(t)+M(t)]/2$.

If one substitutes the asymptotic expansion of $f_{n}(t)$ as a Laurent series, and allows for a change of order between the infinite sum in this expansion and the integral in expression \eqref{tdmtfc}, one gets from that equality a hierarchy of differential equations that allow one to obtain the coefficients of the series for $f_n(t)$ by a recursive process, starting from the dominant contributions in the limit  $\omega_n\rightarrow\infty$ and descending in the sequence of negative integer powers of $\omega_n$. At each step in this hierarchy, new integration constants appear, permitting some freedom in the choice of coefficients for the transformation.

A case in which we are particularly interested is when the transformation does not affect the mass, which becomes then not only mode independent, but besides coincides in the original and the new field descriptions. In this situation, $S(t)$ vanishes, and $\bar{s}(t)=s(t)$. It then follows that $g_{n}(t)=K_{n}-\dot{f}_{n}(t)/2$, where $K_{n}$ is a time constant which depends on the mode. Given the asymptotic behavior of the functions $g_n(t)$ and $f_n(t)$, the constants $K_n$ must have the form
\begin{equation}
K_{n}=\kappa_{0}+\sum_{k=1}^{\infty}\frac{\kappa_{k}}{\omega^{k}_{n}},
\end{equation}
where the $\kappa$'s are real numbers. The expression of the functions $f_n(t)$ can be determined by replacing them with their asymptotic expansion in terms of a Laurent series in Eq. \eqref{tdmtfc}, particularized to the data $S(t)=0$, following then the procedure explained above. As we have commented, solutions can always be found by means of a recursive process, at least at a formal level.

\section{Conclusions}
\label{s5}

We have studied time dependent linear canonical transformations of a certain nonlocal class for scalar fields in (generically) nonstationary spacetimes with compact spatial sections. These time dependent transformations absorb part of the field dynamics in functions of the background. More specifically, the class of transformations that we have studied relate alternate field descriptions in which the equations of motion take the form of a KG equation in an auxiliary, ultrastatic spacetime, but provided with a time dependent mass. We have allowed for subdominant corrections to this field equation in the ultraviolet sector, corresponding to modes with a large eigenvalue (in norm) of the LB operator. The dynamics are completed with the Hamiltonian equation that relates the field configuration and its momentum. We have restricted our discussion to the case in which this latter equation equals the time derivative of the configuration modes with the modes of the momentum, in all of the considered field descriptions. Besides, using the decomposition of the field provided by the eigenfunctions of the LB operator, the possible nonlocality of the analyzed canonical transformations consists in a dependence on the mode under consideration, but respecting the dynamical decoupling between those modes.

For the type of equations of motion that we have permitted, it is known that the criterion of (i) invariance of the vacuum under the spatial symmetries of the dynamics and (ii) existence of a unitary implementation for the evolution, selects a unique class of unitary equivalence for the Fock representations of the CCR's in each of the descriptions related by our canonical transformations. This equivalence class contains the representation determined by the CS which would be natural to associate with the case of a free massless scalar field \cite{cmov-gen,FMOV}. Moreover, the above criterion fixes as well the choice of canonical pair for the field up to \emph{local} linear canonical transformations that vary in time and, when reduced to its action on the field configuration, amount just to a scaling \cite{Crit,FMOV}. Our discussion can be viewed as an extension of these uniqueness results by allowing that (a) the time dependent change in the field configuration includes contributions of the field momentum, and (b) the canonical transformation becomes mode dependent, and hence nonlocal. In this work, we have nonetheless imposed two restrictions on the kind of considered canonical transformations. On the one hand, as we have commented, they must be compatible with the evolution in the sense that they do not mix different modes of the LB operator (which do not interact dynamically). For that, in particular, they must have the same form in each of the eigenspaces of that operator (since the dynamics is the same for all degenerate modes). On the other hand, we have assumed that the functions that determine the canonical transformation admit an asymptotic expansion in the ultraviolet sector in the form of a Laurent series of $\omega_n$ (the square root of the norm of the LB eigenvalue).

We have studied first whether this type of nonlocal and time dependent canonical transformations that connect different descriptions with the same kind of dynamical equations exist or not. We have proven that they do exist in fact, and that they are necessarily mode dependent, so that no local transformation can ever have the required properties. In the two descriptions related by each of these canonical transformations, a privileged Fock quantization is selected by the criterion of unitary evolution and invariance under the spatial symmetries. Then, the question arises of whether these alternate quantizations are in fact equivalent, or whether a new ambiguity appears which affects the physics in the quantum realm, and which is not removed by the proposed criterion. The question can be rephrased by asking whether the considered canonical transformations admit a unitary implementation or not. We have demonstrated that all the studied nonlocal canonical transformations are actually implementable as unitary ones, therefore guaranteeing the uniqueness of the Fock quantum theory picked out by the criterion that we have put forward. This shows the consistency of this criterion and provides further robustness to the selected quantum theory and its physical predictions, beyond the uniqueness results which were already available in the literature \cite{cmov-gen,Crit}. Finally, we have analyzed in more detail the specific case in which the time dependent mass is not corrected by mode dependent subdominant contributions, both in the original and the transformed field descriptions. We have seen that transformations of this type are generally possible, and have formulated the condition that this subfamily of canonical transformations must satisfy.

It is worth noting that this kind of nonlocal transformations appear in cosmological perturbation theory, relating the description of scalar perturbations around FRW closed spacetimes in longitudinal gauge, e.g., and in terms of gauge invariants, as was shown in Ref. \cite{FMOV} and we summarize in the Appendix. To conclude, let us also point out that the results presented in this work seem to admit extensions and further applications in the case of fermionic fields, where this kind of nonlocal transformations may be especially important. For instance, Ref. \cite{Hall} analyzes a quantum model containing fermionic fields with finite particle creation in the evolution or, equivalently, with a unitarily implementable dynamics. To arrive at such a unitary dynamics for the model it is necessary to define suitable annihilation and creationlike variables by using (nonlocal) mode and time dependent canonical transformations with respect to the original description of the fermionic fields. Therefore, our analysis and conclusions may also find important applications in the process of reaching a Fock quantization for fermions with the required good properties of symmetry and unitarity, as well as to ensure the uniqueness (modulo unitary transformations) of such a quantum theory.

\section*{Acknowledgements}

The authors are grateful to L. Castell\'o Gomar, M. Fern\'andez-M\'endez, J. Olmedo, and J. Velhinho for discussions. This work was supported by the research grants MICINN/MINECO FIS2011-30145-C03-02 from Spain and DGAPA-UNAM IN117012-3 from Mexico. D. M-dB acknowledges financial support from CSIC and the European Social Fund under the grant JAEPre\_09\_01796. He also wants to thank the Perimeter Institute for warm hospitality during the preparation of this work.

\appendix
\section{Mode dependent transformations in cosmological perturbations}
\label{a1}

In this appendix, we will consider the application of our discussion to the analysis of scalar perturbations in cosmology \cite{FMOV}. In more detail, the cosmological system that we will discuss is an FRW spacetime with closed spatial sections that have the topology of a three-sphere. The matter content is given by a minimally coupled scalar field of mass $\tilde{m}=m\sqrt{3\pi/2G}$. Starting with a homogeneous setting, one can consider the perturbations around those FRW solutions \cite{hh}, restricting the attention just to scalar perturbations for simplicity \cite{FMOV}. This restriction is possible because the scalar, vector, and tensor perturbations decouple in the evolution at the dominant perturbative order in which one truncates the system \cite{hh}. The expansion of the perturbations in modes is made using hyperspherical harmonics \cite{hh}, which are eigenfunctions of the LB operator on the three-sphere. The nonphysical degrees of freedom can be eliminated, e.g., by adopting a longitudinal gauge. After this gauge fixing, the sector of homogeneous solutions can be described by two canonical pairs, e.g. the pairs $(\alpha,\pi_{\alpha})$ and $(\bar{\varphi},\pi_{\bar{\varphi}})$, used in Ref. \cite{FMOV}, where $\alpha$ is related to the FRW scale factor $a$ and $\bar{\varphi}$ is essentially the homogeneous mode of the scalar matter field \cite{notation}. On the other hand, the scalar perturbations can de described in this gauge fixed model by the set of configuration and momentum pairs $(h_{n\vec{l}},\pi_{n\vec{l}})$, obtained from the mode expansion of the scalar field and its momentum after a scaling of the former by a factor of $a$ (see e.g. Refs. \cite{mukhanov,hh}). The index $n$ is a positive integer such that $n>1$, and labels the eigenspaces of the LB operator on the three-sphere contributing with inhomogeneous physical degrees of freedom to the perturbations, with a corresponding eigenvalue that (after a flip of sign) is equal to $\omega_{n}^{2}=n(n+2)$ \cite{notation}. The degeneration index $\vec{l}$, on the other hand, stands for the pairs of integers $(l,m)$ that designate the different hyperspherical harmonics with the same value of $n$. Their ranges are $0\le l \le n$ and $-l\le m\le l$.

The equations of motion that the inhomogeneous modes satisfy in the longitudinal gauge have the form
\begin{equation}
\label{pieom}
\ddot{h}_{n}+r_{n}(t)\dot{h}_{n}+\left[\omega_{n}^{2}+s_{n}(t)\right]h_{n}=0.
\end{equation}
We obviate the degeneration label $\vec{l}$ because the dynamical evolution is independent of it. The dot stands here for the derivative with respect to the conformal time $t$, and the functions $r_{n}(t)$ and $s_{n}(t)$ are mode and time dependent:
\begin{eqnarray}
\label{irsf}
r_{n}(t)&=&2A_{n}\tilde{g}_{n}^{2}, \\
s_{n}(t)&=&\tilde{s}(t)+\mathcal{O}\left(\omega^{-2}_{n}\right),
\label{irsf2}
\end{eqnarray}
where $\tilde{s}(t)$ is a function that only depends on time, whereas $A_{n}$ and $\tilde{g}^{2}_{n}$ depend also on the considered mode:
\begin{eqnarray}
A_{n}&=&\frac{3}{\omega^{2}_{n}-3}e^{-6\alpha}\pi_{\bar{\varphi}}\left(2\pi_{\alpha}\pi_{\bar{\varphi}}-e^{6\alpha}m^{2}\bar{\varphi}\right), \\ \tilde{g}^{2}_{n}&=&\left[1-\frac{3}{\omega^{2}_{n}-3}e^{-4\alpha}\pi_{\bar{\varphi}}^{2}\right]^{-1}.
\end{eqnarray}
Notice that $A_{n}=\mathcal{O}(\omega^{-2}_{n})$ while $\tilde{g}^{2}_{n}=1+\mathcal{O}(\omega^{-2}_{n})$. It then follows that $r_{n}(t)$ is of the order $\mathcal{O}(\omega_{n}^{-2})$. In Ref. \cite{FMOV}, it was proven that the criterion of spatial symmetry invariance and unitary evolution selects a unique Fock quantization even in this situation, with the KG equation modified by subdominant terms of the above form.\footnote{Actually, the proof is valid for more general types of modifications, like e.g. for corrections to the mass function of order $\mathcal{O}\left(\omega^{-1}_{n}\right)$, as in Eq. \eqref{nmt}.} The Fock representation that is picked out belongs to the unitary equivalence class of that determined by the CS $J_{0}$. It is also possible to obtain an alternate description where the damping term vanishes by considering a mode dependent canonical transformation that eliminates the contribution of $r_n(t)$:
\begin{eqnarray}
\label{1mdct}
\bar{h}_{n}&=&\tilde{g}_{n}h_{n}, \\ \bar{\pi}_{n}&=&\frac{1}{\tilde{g}_{n}}\pi_{n}+\left(\tilde{g}_{n}A_{n}+\dot{\tilde{g}}_{n}\right)h_{n}.
\label{1mdct2}
\end{eqnarray}
Again, we have ignored the degeneration index in the canonical variables, since the transformation is independent of it. With this canonical transformation, $s_{n}(t)$ does not change up to order $\omega^{-2}_{n}$ and hence continues to be given by Eq. \eqref{irsf2}. In addition, a contribution of the configuration modes has been included in the definition of the new momentum modes so that they satisfy the Hamiltonian equation $\bar{\pi}_{n}=\dot{\bar{h}}_{n}$. Therefore, the dynamical equations in this new field description are completely adapted to the form that is considered in the main text of this work.

Alternatively, in the study of cosmological perturbations, it is most common to use gauge invariant quantities to describe the system. A canonical pair of gauge invariant scalar quantities can be constructed from the original variables $(\pi_{n},h_{n})$ by a time dependent linear canonical transformation that depends on the specific mode under consideration, but do not mix those modes:
\begin{eqnarray}
\label{2mdct}
\Psi_{n}&=&\frac{1}{\sqrt{\omega^{2}_{n}-3}}\left(\pi_{n}+\chi h_{n}\right),\\ \Pi_{n}&=&\frac{1}{\sqrt{\omega^{2}_{n}-3}}\left[\chi\pi_{n}+(\chi^{2}-\omega^{2}_{n}+3) h_{n}\right].
\label{2mdct2}
\end{eqnarray} 
We have ignored again the degeneration index on the mode variables, because the transformation has no dependence on it. The function $\chi$ varies in time, and is defined in terms of background variables. It is mode independent, and can be related with $A_{n}$ and $g_{n}$ by the identity:
\begin{eqnarray}\label{cAgr}
\chi&=&\frac{e^{-2\alpha}}{\pi_{\bar{\varphi}}}\left(2\pi_{\alpha} \pi_{\bar{\varphi}}-e^{6\alpha}m^{2}\bar{\varphi}\right)\\
&=&\frac{A_{n}\tilde{g}^{2}_{n}}{1-\tilde{g}^{2}_{n}}.
\end{eqnarray}

The above pair of canonical gauge invariant quantities are linear combinations (depending on the background variables, see Ref. \cite{FMOV} for details) of the energy density and matter velocity perturbations introduced by Bardeen \cite{bardeen}. Their equations of motions are
\begin{eqnarray}
\ddot{\Psi}_{n}+\left[\omega^{2}_{n}+s(t)\right]\Psi_{n}=0,
\end{eqnarray}
 and $\Pi_{n}=\dot{\Psi}_{n}$, up to the perturbation order in which the theory is being truncated. Note that these equations are precisely of the form introduced in Sec. \ref{s2}. Combining the canonical transformation given by Eqs. \eqref{1mdct} and \eqref{1mdct2} with the transformation introduced in Eqs. \eqref{2mdct} and \eqref{2mdct2}, it is easy to obtain the mode and time dependent canonical transformation that relates the gauge invariants $(\Psi_{n},\Pi_{n})$ with the variables $(\bar{h}_{n},\bar{\pi}_{n})$, for all values of $n$. The transformation is
\begin{eqnarray}
\bar{h}_{n}&=&\frac{\tilde{g}_{n}}{\sqrt{\omega^{2}_{n}-3}}\left(-\Pi_{n}+\chi\Psi_{n}\right),
\\ \nonumber
\bar{\pi}_{n}&=&\frac{1}{\sqrt{\omega^{2}_{n}-3}}\left(\chi\tilde{g}_{n}+\dot{\tilde{g}}_{n}\right)\Pi_{n} \\  &\ & \
+\frac{1}{\sqrt{\omega^{2}_{n}-3}}\left(\frac{\omega^{2}_{n}-3}{\tilde{g}_{n}}-\chi^{2}\tilde{g}_{n} -\chi\dot{\tilde{g}}_{n}\right)\Psi_{n}.\qquad
\end{eqnarray}
Using expression \eqref{cAgr}, and in particular that $\chi$ is a mode independent time function, one can check that the functions $f_{n}(t)$ and $g_{n}(t)$ that characterize the mode dependent and time dependent canonical transformation under consideration are
\begin{eqnarray}
f_{n}=-g_{n}\chi &=& -\frac{\tilde{g}_{n}}{\tilde{\omega}_{n}}  \\  &\approx& -\frac{1}{\tilde{\omega}_{n}}-\frac{3}{2\tilde{\omega}^{3}_{n}}e^{-4\alpha}\pi_{\bar{\varphi}}^{2}+\mathcal{O}(\tilde{\omega}^{-5}_{n}). \ \ \
\end{eqnarray}
Here, we have redefined $\tilde{\omega}^{2}_{n}=\omega^{2}_{n}-3$. In the light of the analysis performed in Sec. \ref{s3-3}, this transformation is of the type given in expression \eqref{pt2}, where the coefficients of the linear combinations that provide the new configuration modes vanish in the ultraviolet limit, while the coefficients of the new momentum modes blow up. As a corollary, the transformation admits a unitary implementation in the Fock representation determined by $J_0$ in the gauge invariant description, and hence the quantization reached in the longitudinal gauge is unitarily equivalent.


\begin{thebibliography}{99}
\bibitem{notation} Compared to Ref. \cite{FMOV}, we use here a slightly different notation.
\end{thebibliography}
\end{document}